\documentstyle[aps,prl,preprint,floats,epsfig]{revtex}

\long\def\onefig#1#2#3
{
\begin{figure}
   \centering \leavevmode
   \epsfxsize=7.0in
   \epsfbox{#1}
   \vskip -0.15in
   \caption{\normalsize \label{fig:#2} #3}
   \hfill
\end{figure}
}
\long\def\twofig#1#2#3#4{ 
\begin{figure}[htbp]
\vspace{9cm}
        \begin{minipage}[t]{8cm}
                \includegraphics{#1}
        \end{minipage}
        \hfill
        \begin{minipage}[t]{8cm}
                \includegraphics{#2}
        \end{minipage}
	\vskip 2cm
        \caption{ \label{fig:#3} #4 }
\end{figure}
}

\begin{document}

\preprint{\tighten\vbox{\hbox{\hfil SMUHEP-00-03}}}

%\draft

% Comment out to get double spacing.
\tighten

\title{Expected Coverage of Bayesian Confidence Intervals for the Mean of
       a Poisson Statistic in Measurements with Background}  

\author{Ilya Narsky
\thanks{Tel.: 1 607 254 2778; Fax: 1 607 255 8062; E-mail:
narsky@mail.lns.cornell.edu;
Mail: Wilson Laboratory, SMU Group, Ithaca, NY, 14853, USA.}} 
\address{Physics Department, Southern Methodist University, Dallas,
TX, 75275-0175, USA}

\date{\today}
\maketitle

\begin{abstract} 
Expected coverage and expected length of 90\% upper and lower limit
and 68.27\% central intervals are plotted as functions of the true
signal for various values of expected background. Results for several
objective priors are shown, and formulas for calculation of confidence
intervals are obtained. For comparison, expected coverage and length
of frequentist intervals constructed with the unified approach of
Feldman and Cousins and a simple classical method are also shown. It is
assumed that the expected background is accurately known prior to
performing an experiment. 
The plots of expected coverage and length are provided for values
of signal and background typical for particle experiments with
small numbers of events.

\end{abstract}


\section{Introduction}

An intense discussion of various methods for setting confidence limits
has led to a number of recent publications in the particle physics
community. Many such papers discuss methods for confidence interval 
construction in experiments with small numbers of signal events.
The outcome of such an experiment is modeled by a Poisson statistic
\begin{equation}
\label{eq:poiss}
f(n|s) = e^{-(s+b)} (s+b)^n/n!\ ,
\end{equation}
where $f(n|s)$ is the conditional probability of observing $n$ events
given the signal $s$. It is assumed in this note that the expected
background $b$ is accurately known before performing an experiment and
can be treated as constant.

Expected coverage and interval length are important properties of a method.
These properties were studied in Refs.~\cite{innocente,helene1,zech}, 
but there the discussion was mostly focused on Bayesian methods with 
a uniform prior. In this paper I compare expected coverage and interval length 
for several objective Bayesian methods and the unified approach~\cite{unified}
of Feldman and Cousins.

The expected coverage for a specific method is defined as
\begin{eqnarray}
\label{eq:cov}
C(s)   & = & \sum_{n=0}^{\infty} I(n,s) f(n|s)\ ; \\
I(n,s) & = & \left\{
\begin{array}{cl}
1 & \mbox{if $s_1(n) \leq s \leq s_2(n)$}\ ; \\
0 & \mbox{otherwise}.
\end{array}
\right. \nonumber
\end{eqnarray}
where $[s_1(n),s_2(n)]$ is the confidence interval constructed for the
number of events $n$ with this method. 
The expected length is given by
\begin{equation}
\label{eq:len}
L(s) = \sum_{n=0}^{\infty} \left(s_2(n)-s_1(n)\right) f(n|s)\ .
\end{equation}
For lower limit intervals $[s_1,+\infty)$ the expected length is infinite,
and instead of the length~(\ref{eq:len}) an expected distance
from zero to the lower limit is plotted:
\begin{equation}
\label{eq:len1}
L_1(s) = \sum_{n=0}^{\infty} s_1(n) f(n|s)\ .
\end{equation}

Coverage is mostly a frequentist concept. In a Bayesian analysis one
is not concerned about coverage as much as about adequately representing
one's objective or subjective prior belief. This, of course, should not
prevent us from investigating properties of expected coverage for various
Bayesian methods. Another important question is how one should define the
desired coverage. Due to the discreteness of the Poisson pdf~(\ref{eq:poiss}),
none of the methods provides exact coverage for all values of $s$. 
In a frequentist approach one typically requires at least minimal coverage
for every value of the true signal $s$, and thus frequentist intervals
overcover on the average. This overcoverage is not an intrinsic feature
of the frequentist interval construction but merely a consequence of the
empirical conservatism that many physicists find desirable. Statistics
literature has examples of consistent frequentist 
methods~\cite{wilson,agresti} 
that lack this minimal coverage property. One can argue that a method 
that gives the requested mean coverage should be preferred over the one that 
always overcovers. No matter what the solution to this problem is, further
discussion of this issue is beyond the scope of this paper.

While confidence intervals and the expected coverage are invariant
under metric transformation, the expected length and mean coverage
are not. A transformation of metric, of course, involves a corresponding 
Jacobian transformation of the prior density in a Bayesian approach.
For example, if the confidence interval for $s$ is $[s_1(n),s_2(n)]$, then
a confidence interval for the quantity $1/s$ is obtained by a simple 
inversion: $[1/s_2(n),1/s_1(n)]$, and the expected coverage has the same
functional dependence upon signal $s$. But the expected interval length
and mean coverage apparently change due to the choice of a new metric.

\section{Bayesian Intervals}

In a Bayesian approach one has to assume a prior probability
density function $\pi(s)$ for unknown signal and then obtain a posterior pdf
\begin{equation}
\label{eq:poster}
\pi(s|n) = \frac{ f(n|s) \pi(s) }{ \int_{0}^{\infty} f(n|s) \pi(s) ds }\ .
\end{equation}
A confidence interval $[s_1,s_2]$ is then found by solving equations
\begin{eqnarray}
\label{eq:bayci}
\left\{
\begin{array}{ccr}
\alpha_1 & = & \int_{0}^{s_1} \pi(s|n) ds\ ; \\
\alpha_2 & = & \int_{s_2}^{+\infty} \pi(s|n) ds\ ; 
\end{array}
\right. 
\end{eqnarray}
for $s_1$ and $s_2$. Here, $\alpha_1$ and $\alpha_2$ are the probabilities 
of the left and right tails of the posterior pdf, respectively, and
\begin{equation}
\mbox{CL}=1-\alpha_1-\alpha_2
\end{equation}
is the confidence level of the constructed interval $[s_1,s_2]$.
For example, to compute a 90\% upper limit, one should put $\alpha_1=0$ and
$\alpha_2=0.1$; for a 90\% lower limit these values are $\alpha_1=0.1$
and $\alpha_2=0$; and to obtain a 68.27\% central interval, one should
use $\alpha_1=\alpha_2=0.15865$.

In particle physics it has been customary to choose a ``non-informative''
or ``objective'' prior $\pi(s)$, i.e., a prior that claims absence
of any knowledge about the true signal $s$. The three most popular candidates 
for an objective prior in a measurement without background are 
a flat pdf~\cite{bayes,laplace,helene2}, 
Jeffreys' prior $1/s$~\cite{jeffr1,jeffr2,jaynes} and
the $1/\sqrt{s}$ prior~\cite{box}.
The last two were derived from first principles, and the motivation for
a flat prior is merely that it is ``obvious''.
If non-zero background is expected, one can apply the same first principles
and obtain $1/(s+b)$ and $1/\sqrt{s+b}$ for the two derived prior pdf's,
respectively. One can also argue~\cite{linnemann} that the $1/\sqrt{s}$ prior 
should be used even if non-zero background is expected 
because our prior knowledge of background should
not affect our prior ignorance about signal. The latter argument can be
applied to the $1/s$ prior as well, but this prior gives a divergent posterior
pdf for non-zero background $b>0$ and is therefore unacceptable. 
The discussion below is confined to these four prior pdf's: flat, $1/(s+b)$,
$1/\sqrt{s}$ and $1/\sqrt{s+b}$.

For a prior distribution
\begin{equation}
\label{eq:bay1}
\pi(s) = 1/s^m\ ;\ \ \ \ 0 \leq m < 1\ ;
\end{equation}
the posterior pdf is given by
\begin{equation}
\label{eq:post1}
\pi(s|n) = 
\frac{ e^{-s} s^{-m} (s+b)^n/n! }
{ \sum_{k=0}^{n} b^k \Gamma(n-k-m+1)/\left(k!(n-k)!\right) }\ ,
\end{equation}
and the confidence interval can be computed using
\begin{eqnarray}
\label{eq:int1}
\left\{
\begin{array}{ccr}
\alpha_1 & = & 1 - 
\left( \sum_{k=0}^{n} b^k \frac{\Gamma(n-k-m+1,s_1)}{k!(n-k)!} \right) /
\left( \sum_{k=0}^{n} b^k \frac{\Gamma(n-k-m+1)}{k!(n-k)!} \right)\ ; \\
\alpha_2 & = & 
\left( \sum_{k=0}^{n} b^k \frac{\Gamma(n-k-m+1,s_2)}{k!(n-k)!} \right) /
\left( \sum_{k=0}^{n} b^k \frac{\Gamma(n-k-m+1)}{k!(n-k)!} \right)\ ; 
\end{array}
\right. 
\end{eqnarray}
where  
\begin{equation}
\Gamma(p,\mu) = \int_{\mu}^{\infty} x^{p-1} e^{-x} dx ;
\ \ p>0 ;\ \ \mu \geq 0 ;
\end{equation}
is an incomplete gamma-function, and
$\Gamma(p)=\Gamma(p,0)$ is the standard gamma-function.

For a prior distribution
\begin{equation}
\label{eq:bay2}
\pi(s) = 1/(s+b)^m\ ;\ \ \ \ 0 \leq m \leq 1\ ;
\end{equation}
a similar derivation leads to the posterior pdf
\begin{equation}
\label{eq:post2}
\pi(s|n) = \frac{ e^{-(s+b)} (s+b)^{n-m} }{ \Gamma(n-m+1,b) }\ .
\end{equation}
The confidence interval is then given by
\begin{eqnarray}
\label{eq:int2}
\left\{
\begin{array}{ccr}
\alpha_1 & = & 1 - 
\frac{ \Gamma(n-m+1,s_1+b) }{ \Gamma(n-m+1,b) }\ ; \\
\alpha_2 & = & 
\frac{ \Gamma(n-m+1,s_2+b) }{ \Gamma(n-m+1,b) }\ . 
\end{array}
\right. 
\end{eqnarray}
At $m=0$ (flat prior) or $b=0$ (no background) 
the posterior distributions~(\ref{eq:post1}) 
and (\ref{eq:post2}) are, of course, identical. 
The posterior pdf~(\ref{eq:post2}) is divergent 
at $m=1$ (Jeffreys' prior) and $n=0$ (no events observed).
I assume that this is equivalent to producing an interval of zero length
and set $I(0,s)$ in Eqn.~(\ref{eq:cov}) to zero.

\section{Frequentist Intervals}

For comparison, expected coverage and length of confidence intervals
constructed with the unified approach~\cite{unified} of Feldman and Cousins
and a simple classical method are also plotted. The simple classical
approach, to which I will refer below as ``standard'', assumes a confidence
interval of the form $[0,s_2]$ for upper limit calculation and that
of the form $[s_1,+\infty)$ for a lower limit. Rules for construction
of frequentist confidence belts are outlined in a famous paper~\cite{neyman} 
by Neyman. The confidence belt $[n_1(s),n_2(s)]$
for the standard classical method is given by
\begin{eqnarray}
\label{eq:stand}
\left\{
\begin{array}{ccr}
n_1 & = & \mbox{smallest integer for which 
               $\sum_{k=0}^{n_1} f(k|s) > \alpha_2$}\ ; \\
n_2 & = & \mbox{largest integer for which 
               $\sum_{k=n_2}^{\infty} f(k|s) > \alpha_1$}\ .
\end{array}
\right. 
\end{eqnarray}
Here, the same convention as in the previous section is used for
$\alpha_1$ and $\alpha_2$, e.g., to compute a 90\% upper limit, one should
put $\alpha_1=0$ and $\alpha_2=0.1$ etc.
This belt is then used to construct a confidence interval $[s_1(n),s_2(n)]$
for a specific number of events $n$ in accordance with Ref.~\cite{neyman}.

\section{Discussion}

Expected coverage and interval length for the four objective priors
and two frequentist methods are plotted versus true signal in the
range $0\leq s\leq 10$ with a step $0.001$ for 4 different values of 
expected background $b=0,\ 1,\ 3,\ \mbox{and}\ 6$. The plots are shown in 
Figs.~1-12. The expected length was computed by summing terms in
Eqn.~(\ref{eq:len}) from 0 to 100 which was enough to achieve a good
accuracy for the chosen values of $s$ and $b$. 

Both frequentist methods produce intervals of at least minimal coverage
because this conservative requirement was used in construction of their
confidence belts. Without background the Bayesian method with a flat prior 
and the standard classical procedure give identical upper limit values, and 
this is also true for the Bayesian with Jeffreys' prior and the
standard classical methods in case of lower limits. At any background 
the Bayesian method with a flat prior provides at least minimal coverage for 
upper limit intervals and undercovers for lower limits, and this is reversed 
for Jeffreys' prior. None of the Bayesian procedures
provides minimal coverage for all values of the true signal and for all
types of confidence intervals. In terms of coverage, the $1/\sqrt{s+b}$ prior
is the most versatile choice among the Bayesian methods. 
It provides a reasonable mean coverage for all types
of confidence intervals while the $1/\sqrt{s}$ prior tends to
undercover for upper limits and overcover for lower limits. 
For central intervals the expected coverage of all Bayesian methods oscillates
about the required confidence level. 
The Bayesian method with a flat prior always gives longer 
central and upper limit intervals than that with the $1/\sqrt{s+b}$ prior,
and intervals produced with this prior are in turn longer than 
those obtained with Jeffreys' $1/(s+b)$ prior. The $1/\sqrt{s}$ prior
produces short intervals at small signal values but, as the signal increases, 
these intervals become longer than those produced 
by the $1/\sqrt{s+b}$ and $1/(s+b)$ priors. 
Expected lengths for all Bayesian methods approach each other asymptotically
as the signal increases. The unified approach produces short 90\%
intervals at large signal values, but one should keep in mind that here
the expected length of a two-sided unified interval is plotted as opposed
to lengths of one-sided intervals for all other methods.

\section{Acknowledgements}

Thanks to Robert Cousins for useful discussion and reading the draft of 
this paper. Thanks to Jim Berger for useful comments.

\onefig{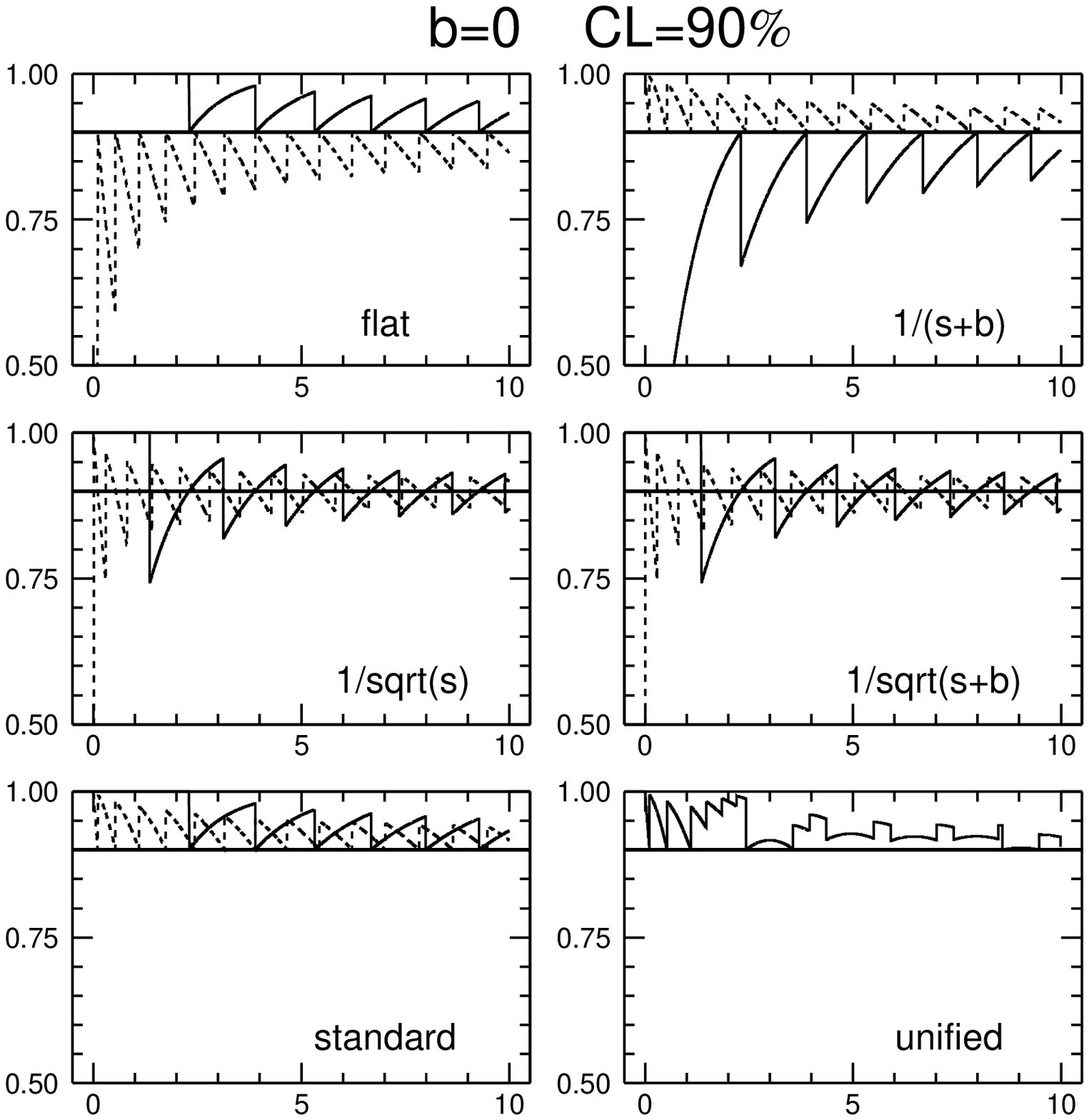}{b0_90}{Expected coverage of 90\% upper limit (solid) 
and lower limit (dashed) intervals for the expected background $b=0$. 
For the unified approach the expected coverage of 90\% intervals is plotted.}

\onefig{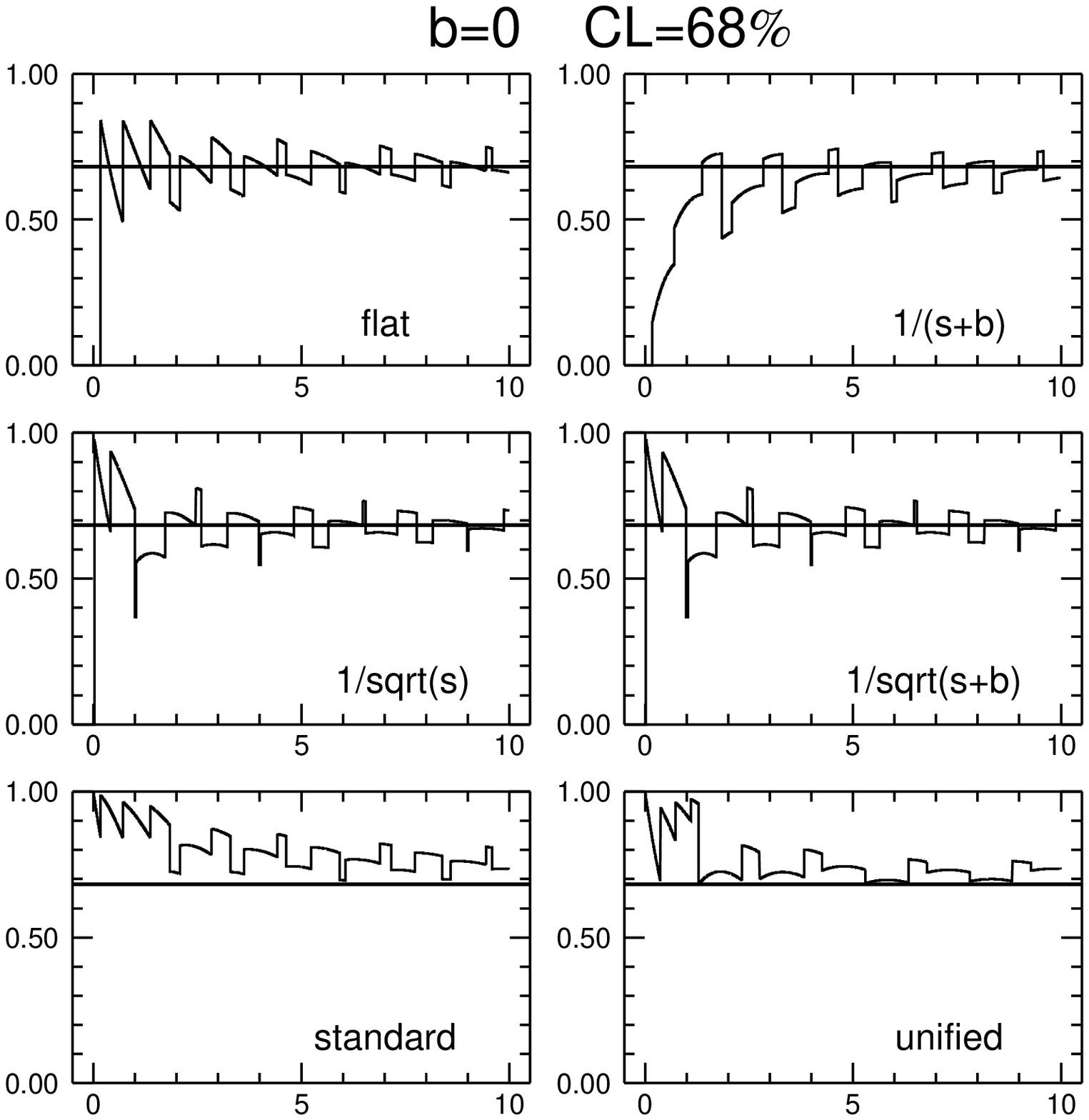}{b0_68}{Expected coverage of 68.27\% central intervals 
for the expected background $b=0$.}

\onefig{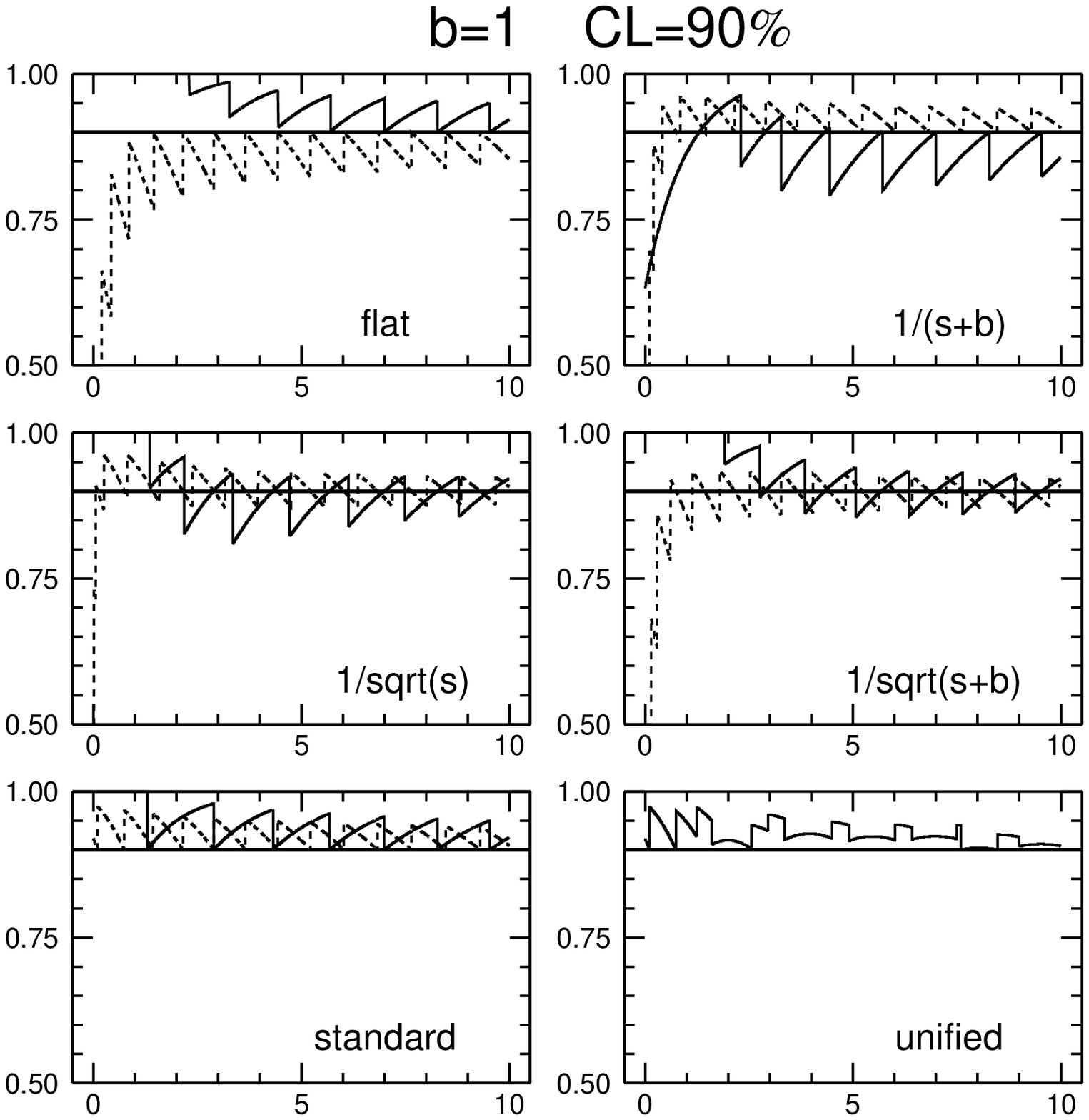}{b1_90}{Expected coverage of 90\% upper limit (solid) 
and lower limit (dashed) intervals for the expected background $b=1$. 
For the unified approach the expected coverage of 90\% intervals is plotted.}

\onefig{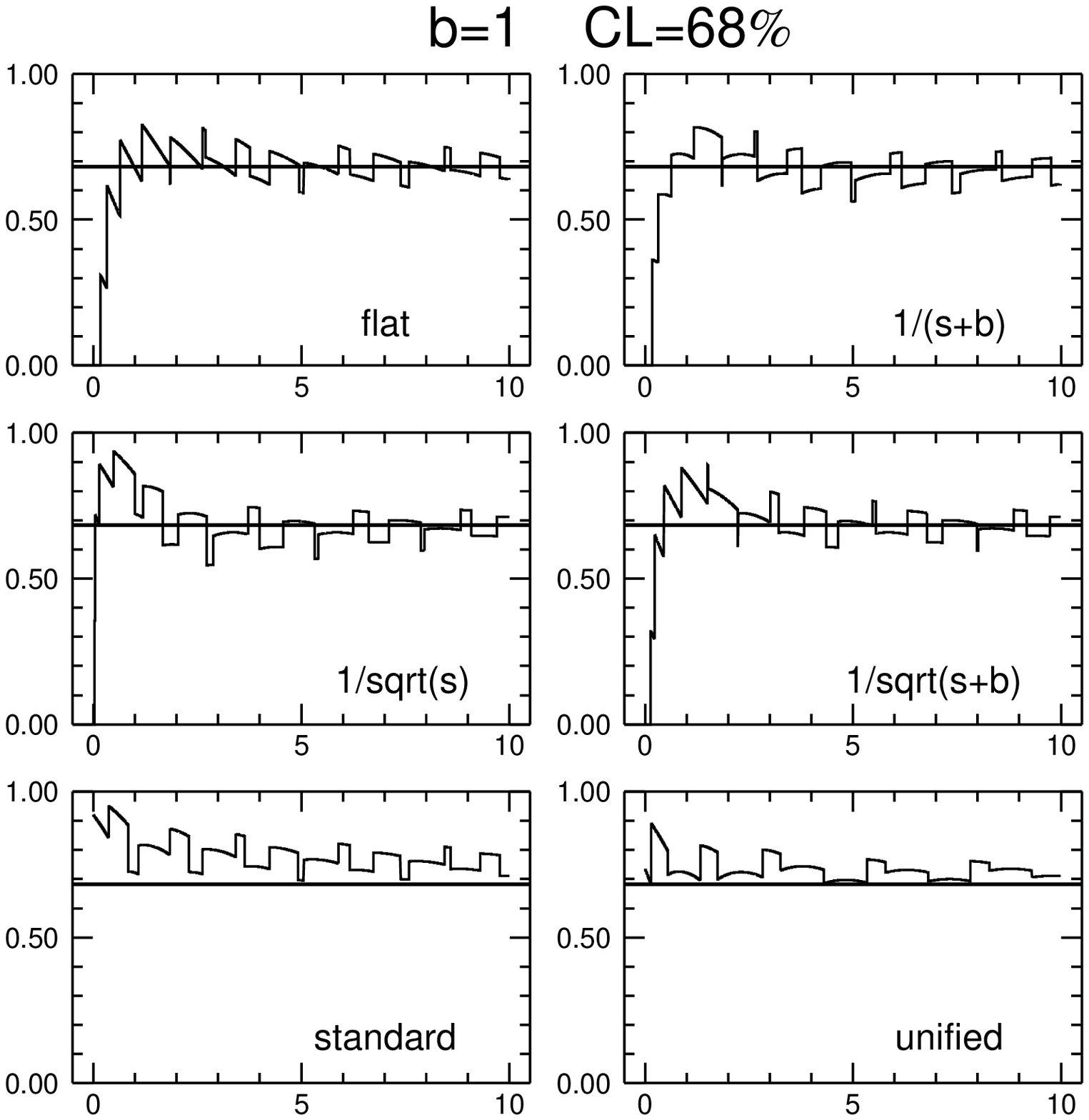}{b1_68}{Expected coverage of 68.27\% central intervals 
for the expected background $b=1$.}

\onefig{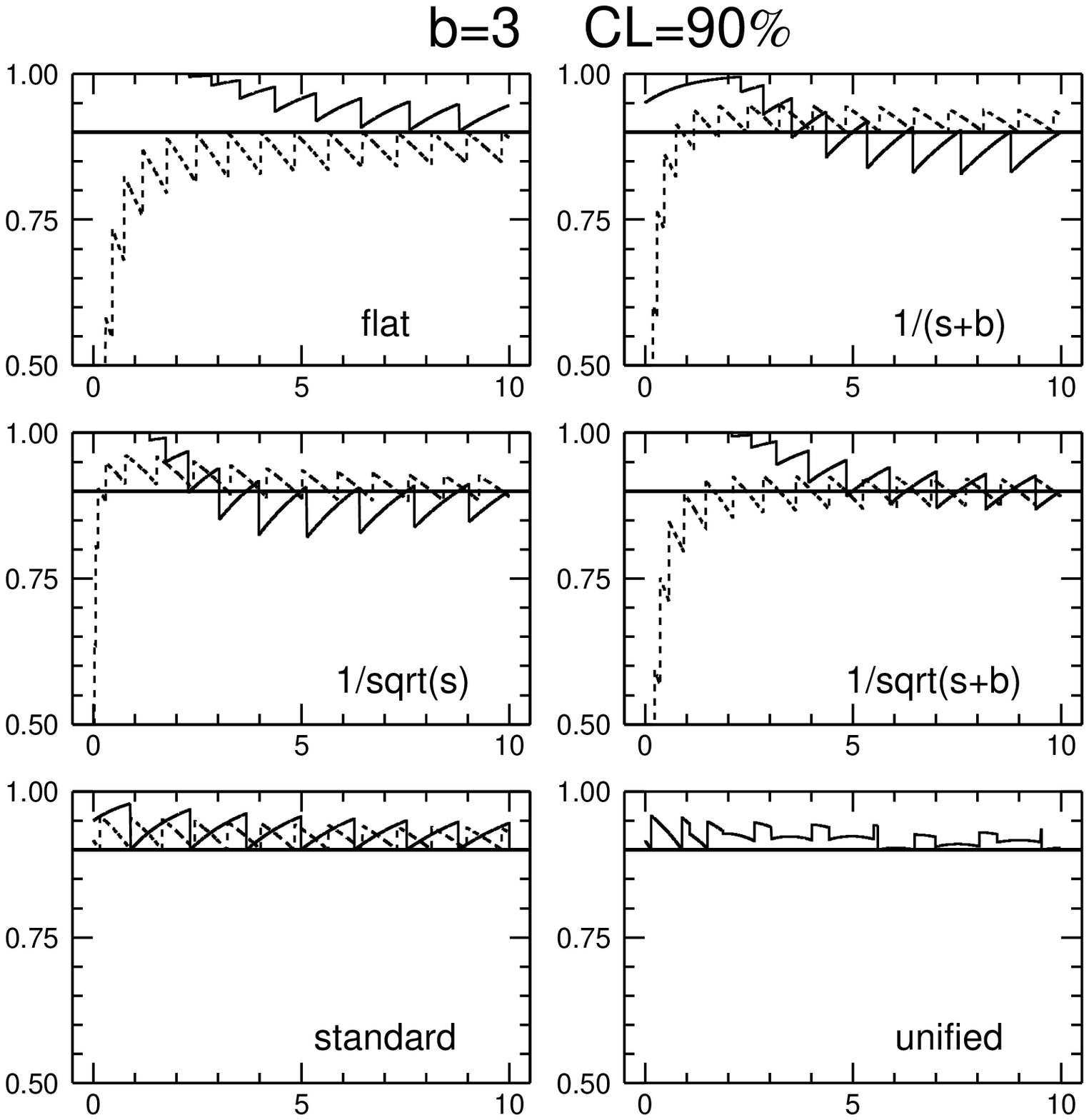}{b3_90}{Expected coverage of 90\% upper limit (solid) 
and lower limit (dashed) intervals for the expected background $b=3$. 
For the unified approach the expected coverage of 90\% intervals is plotted.}

\onefig{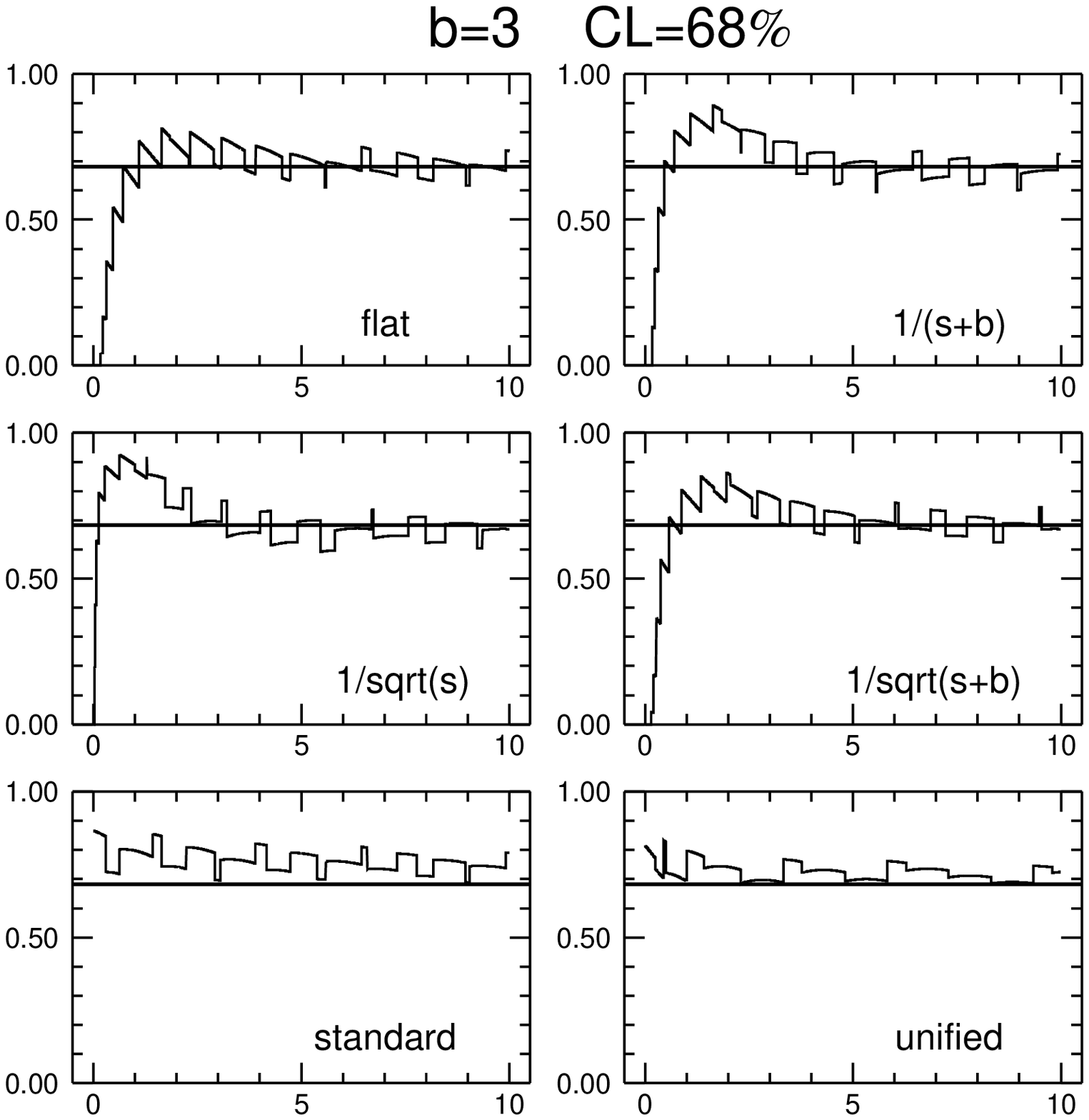}{b3_68}{Expected coverage of 68.27\% central intervals 
for the expected background $b=3$.}

\onefig{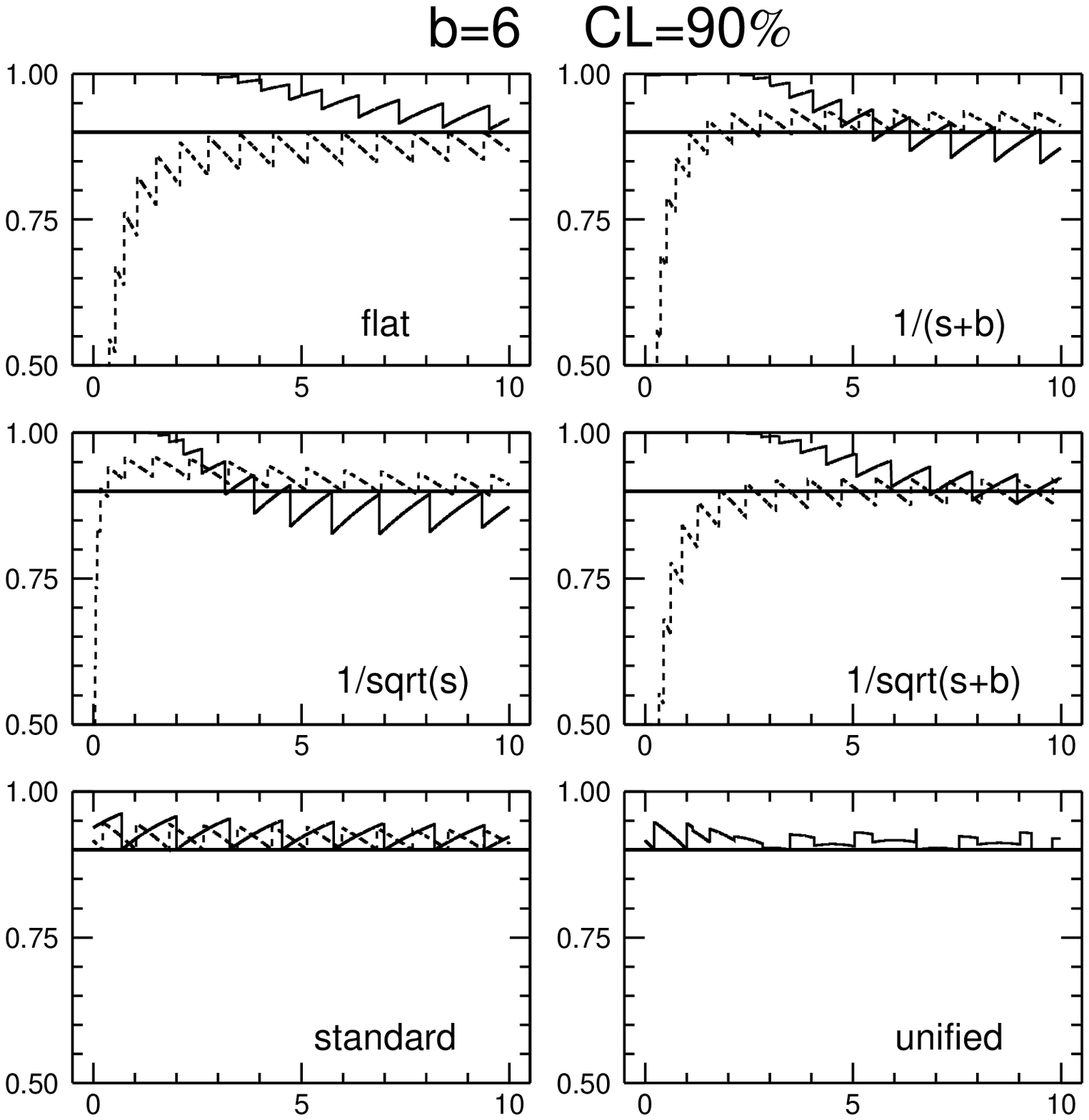}{b6_90}{Expected coverage of 90\% upper limit (solid) 
and lower limit (dashed) intervals for the expected background $b=6$. 
For the unified approach the expected coverage of 90\% intervals is plotted.}

\onefig{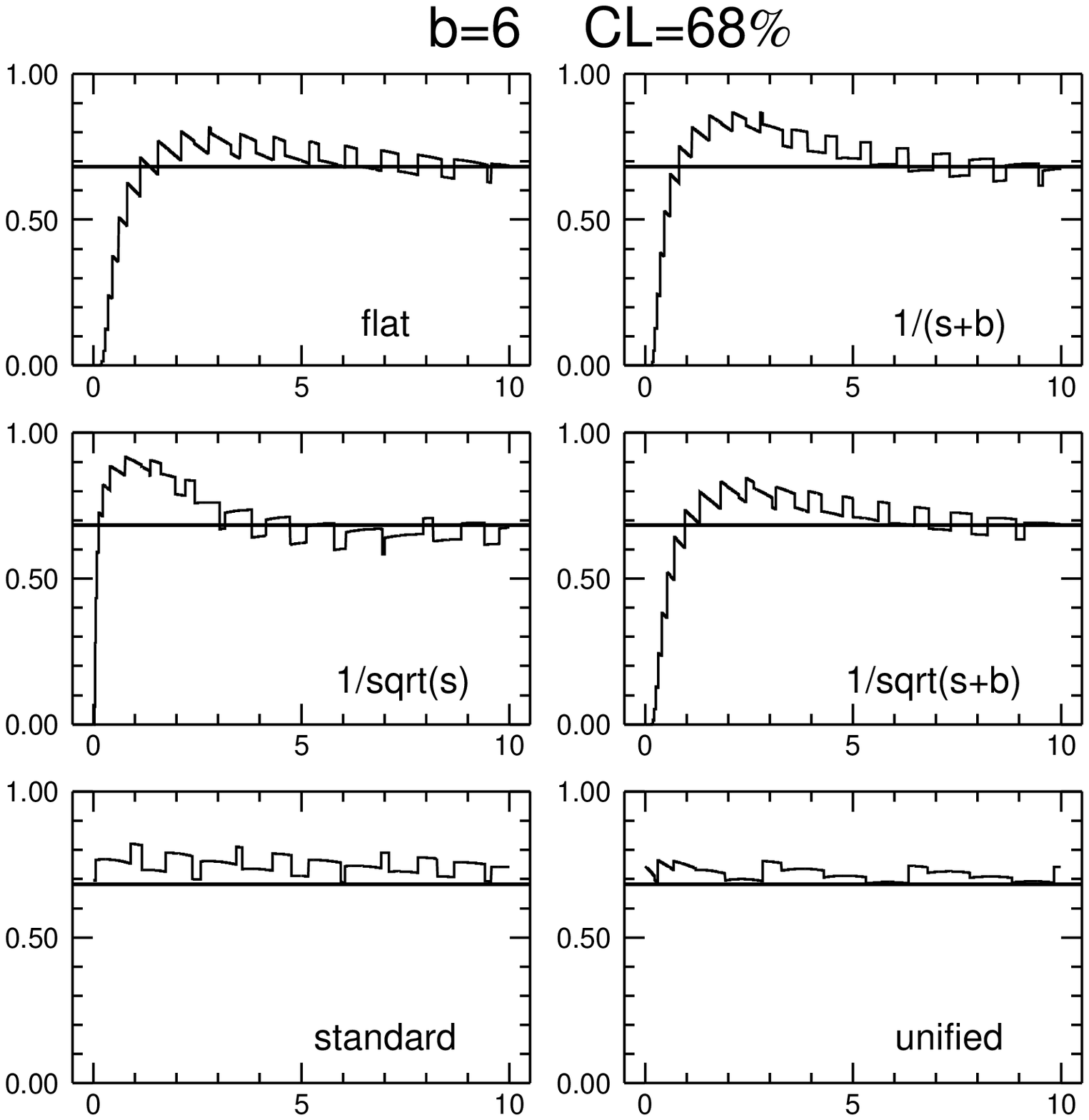}{b6_68}{Expected coverage of 68.27\% central intervals 
for the expected background $b=6$.}

\twofig{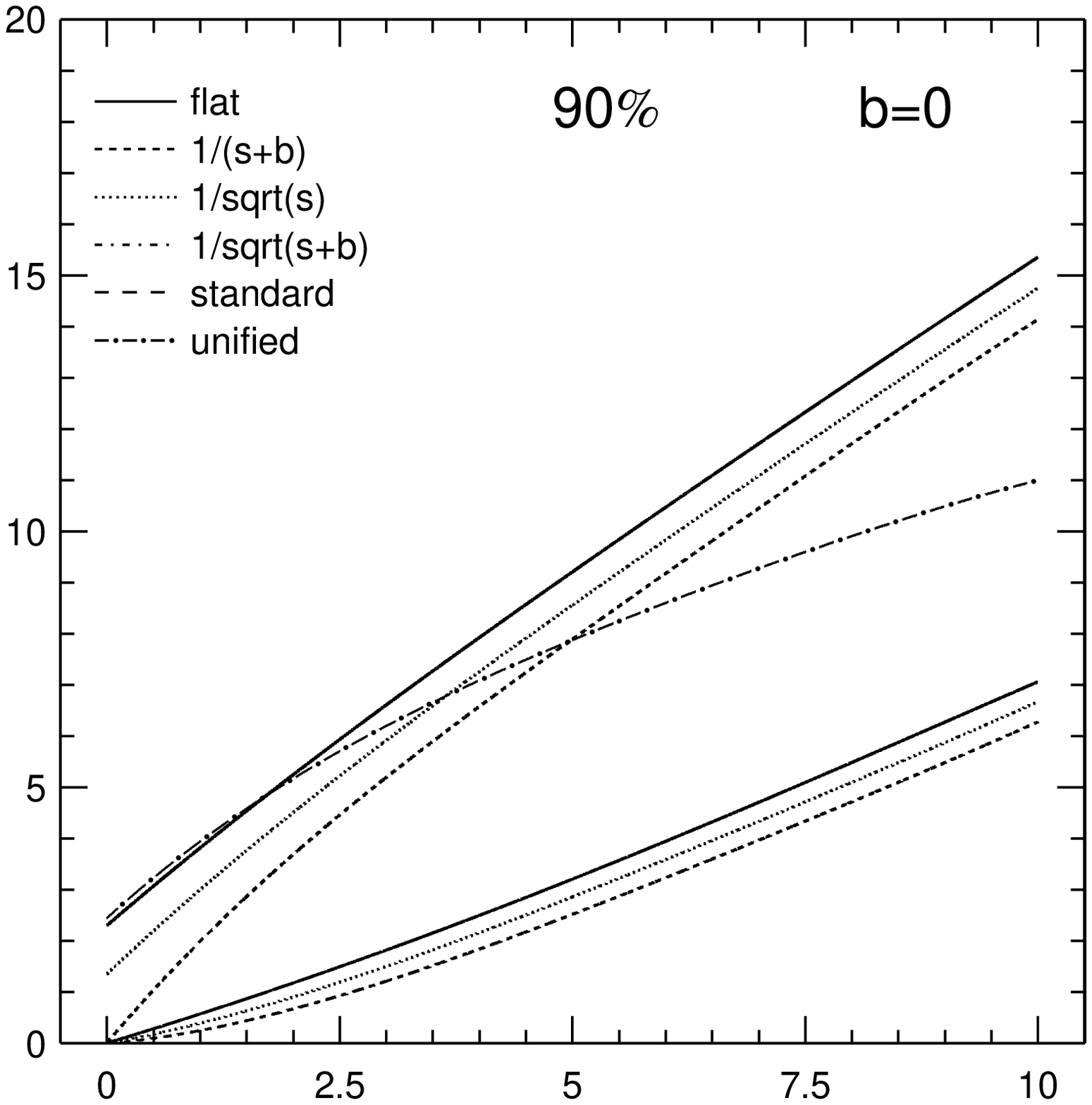}{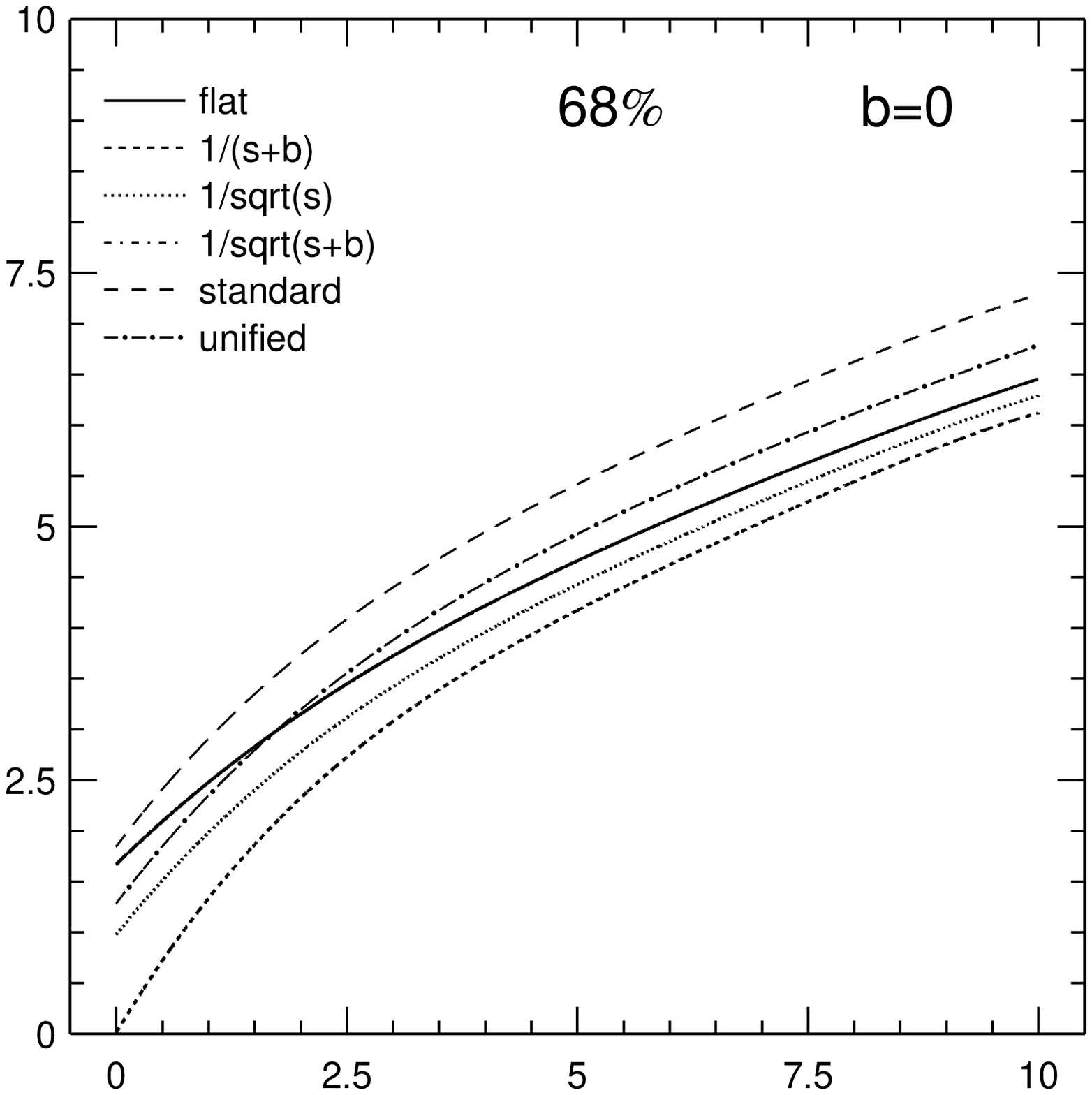}{b0_len}
{Expected length of 90\% upper limit (upper bunch of curves) 
and lower limit (lower bunch of curves) intervals on the left
and expected length of 68.27\% central intervals on the right.
The expected background is $b=0$.}

\twofig{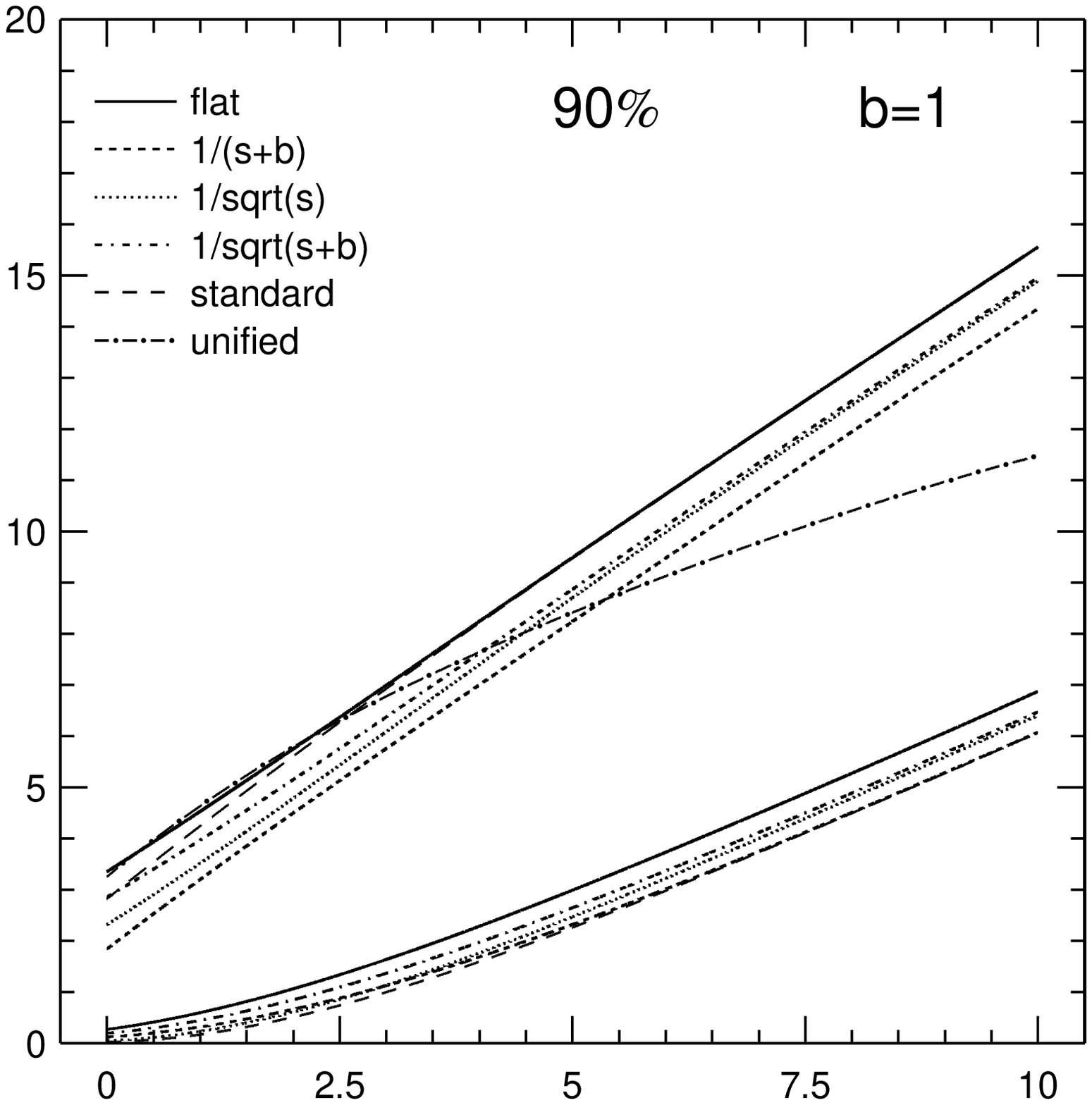}{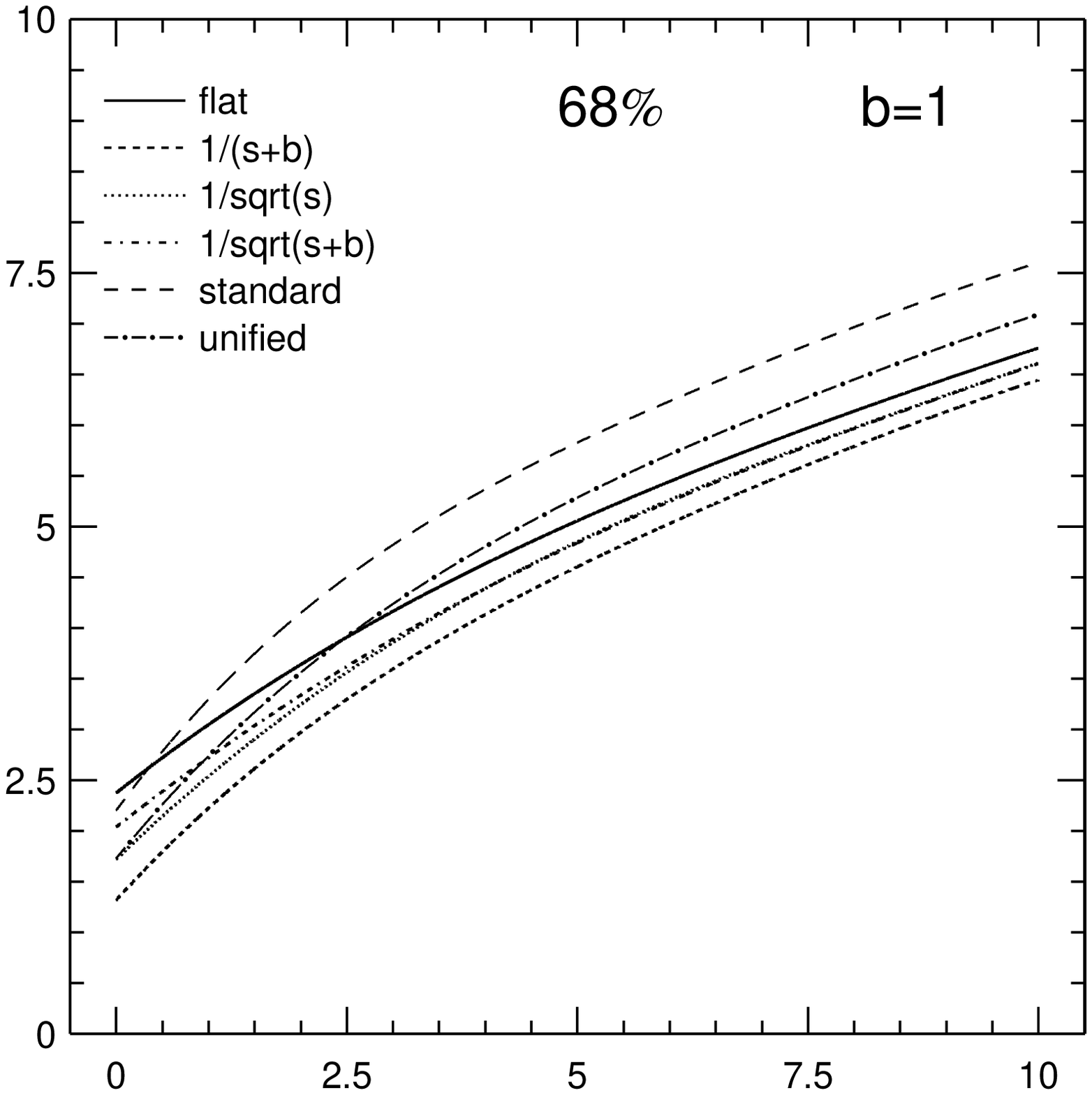}{b1_len}
{Expected length of 90\% upper limit (upper bunch of curves) 
and lower limit (lower bunch of curves) intervals on the left
and expected length of 68.27\% central intervals on the right.
The expected background is $b=1$.}

\twofig{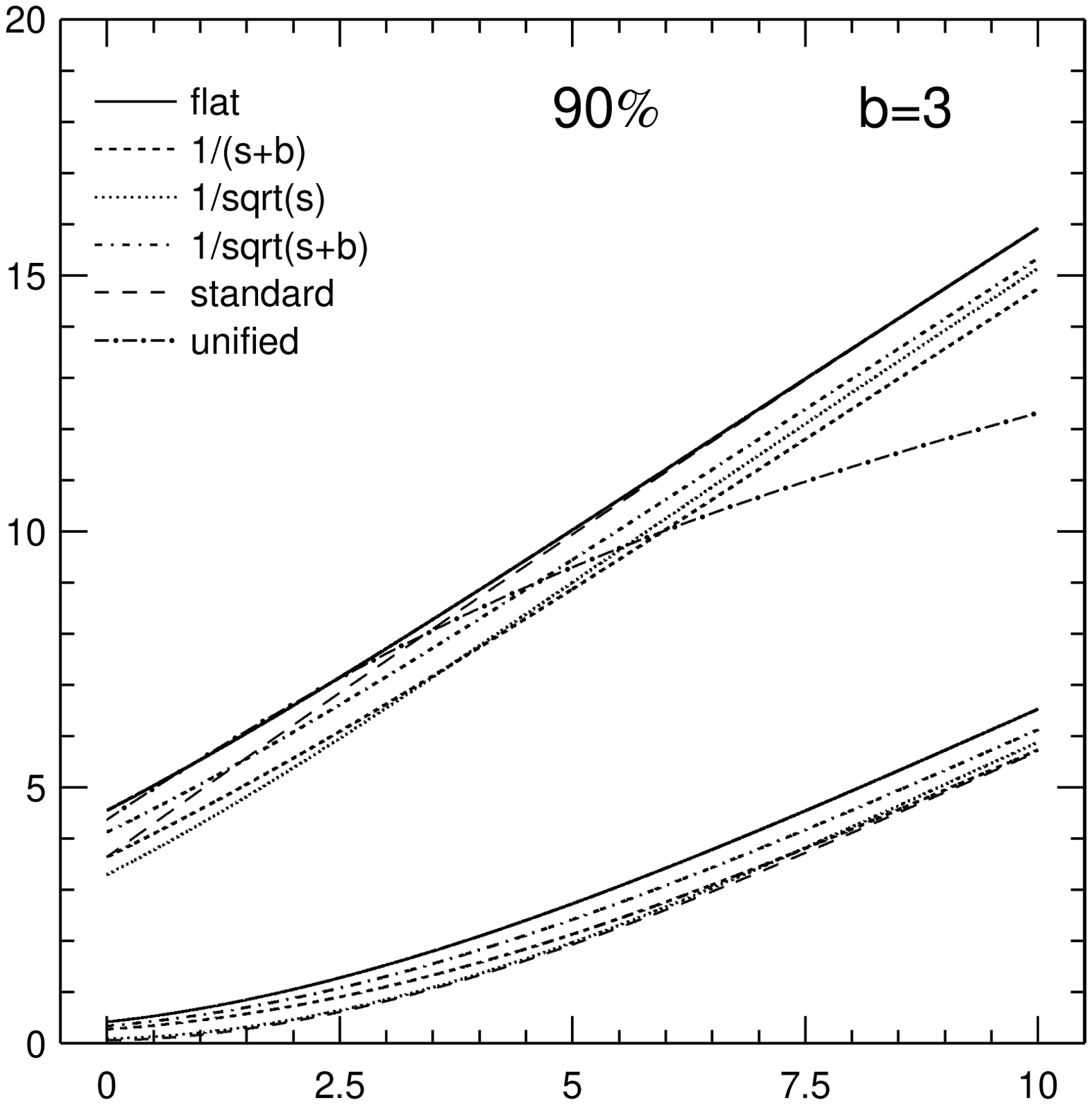}{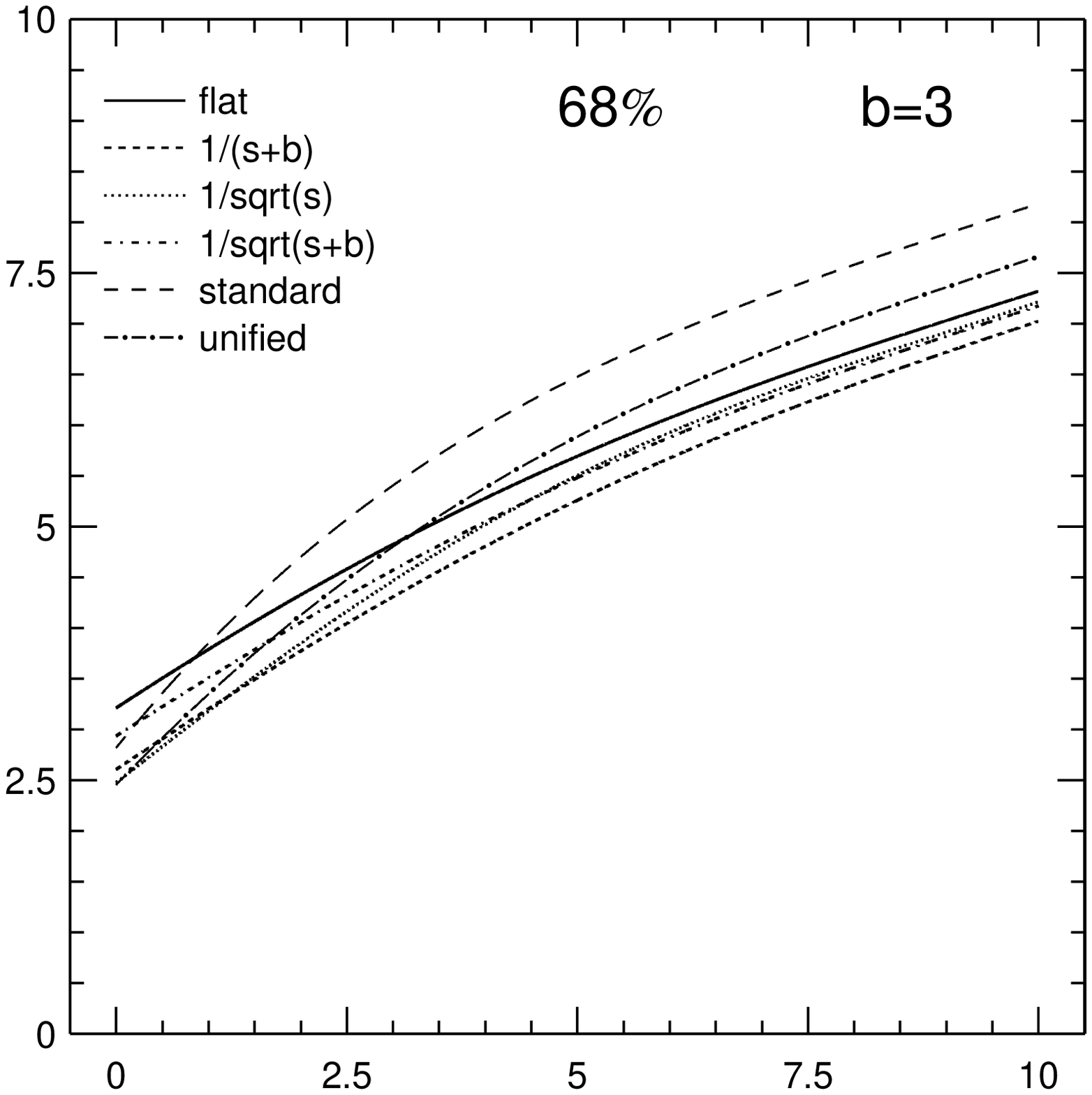}{b3_len}
{Expected length of 90\% upper limit (upper bunch of curves) 
and lower limit (lower bunch of curves) intervals on the left
and expected length of 68.27\% central intervals on the right.
The expected background is $b=3$.}

\twofig{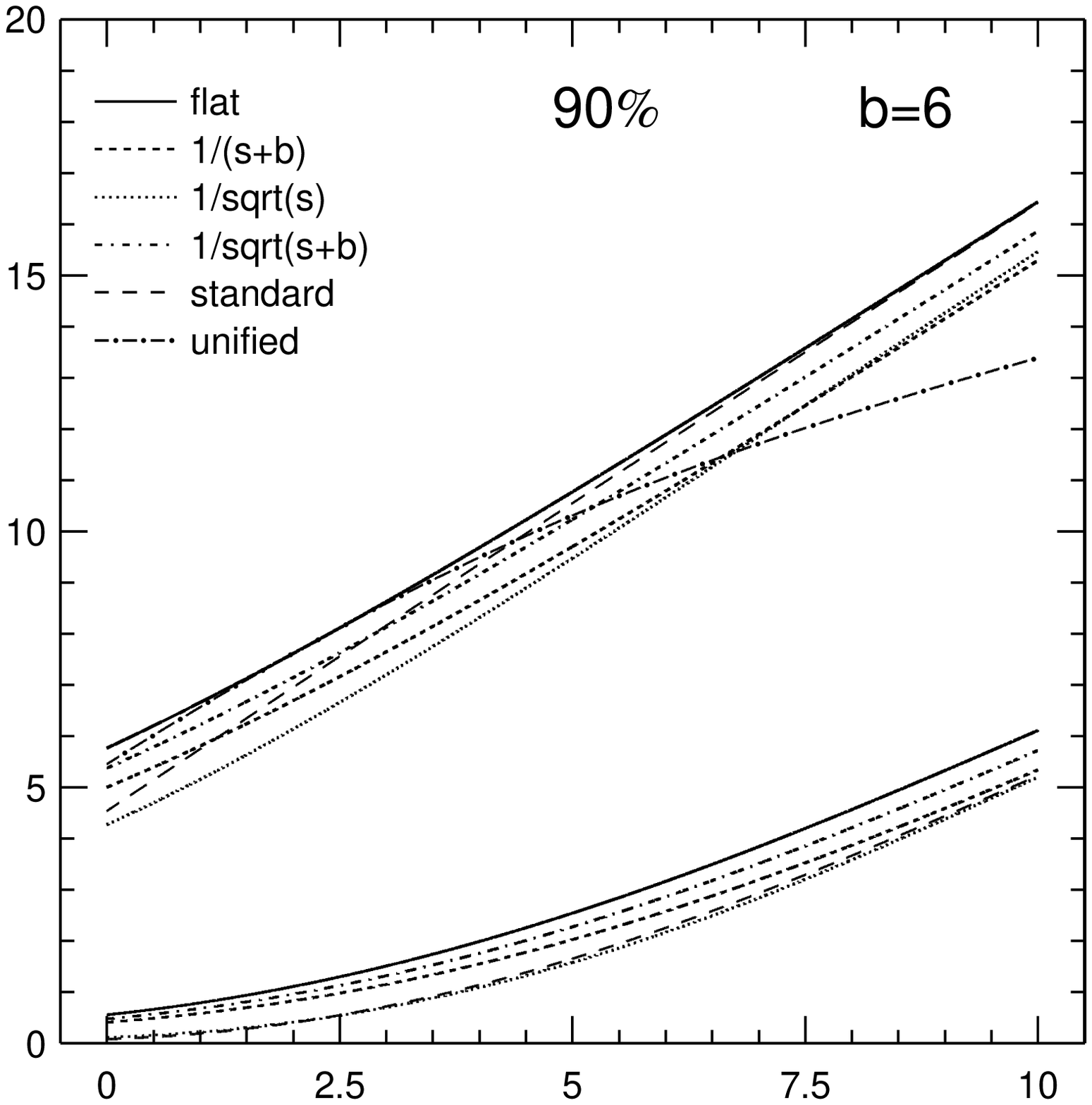}{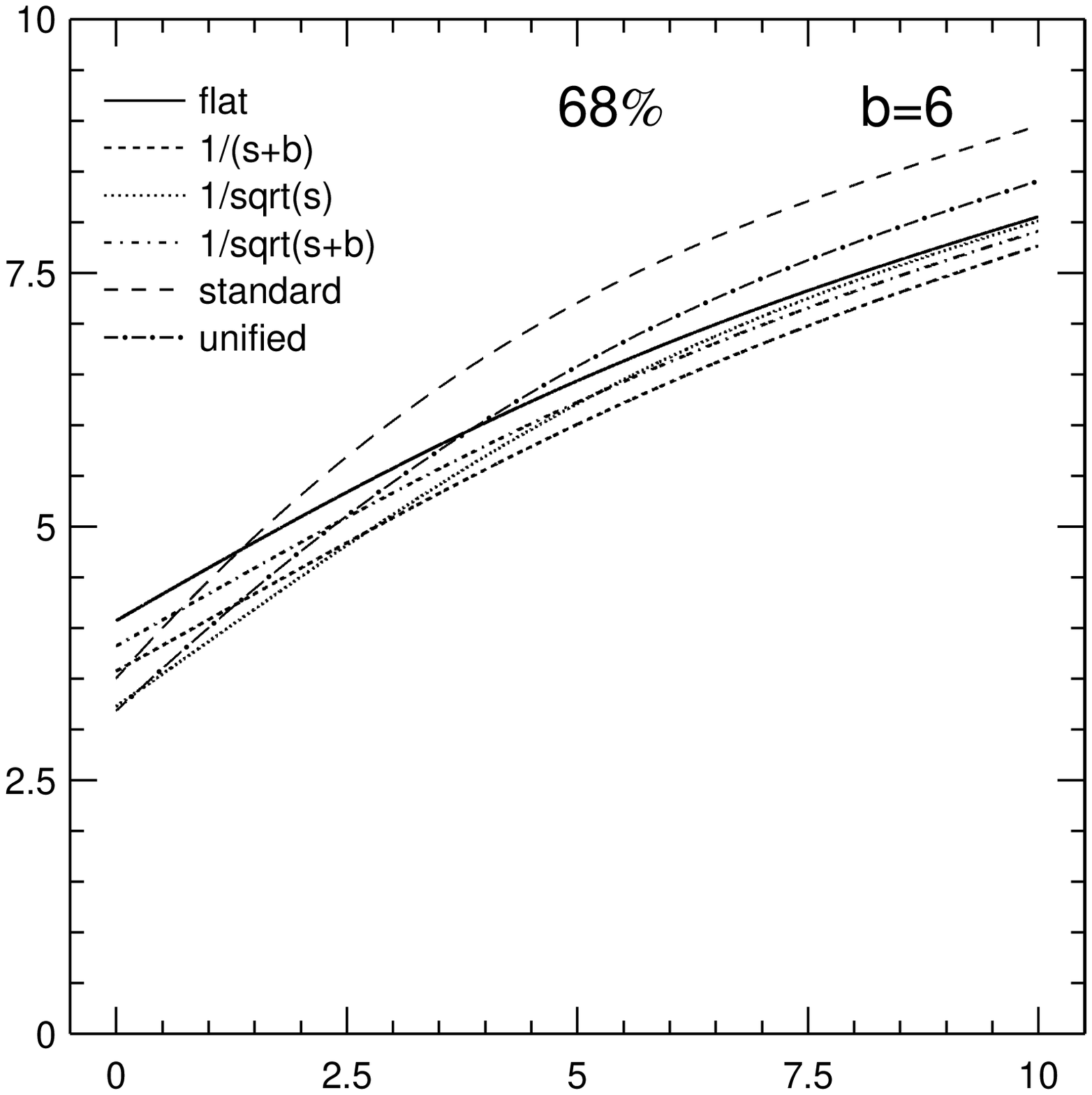}{b6_len}
{Expected length of 90\% upper limit (upper bunch of curves) 
and lower limit (lower bunch of curves) intervals on the left
and expected length of 68.27\% central intervals on the right.
The expected background is $b=6$.}

\end{document}